# Model Specification in Mixed-Effects Models: A Focus on Random Effects (v3.0)

Keith R. Lohse, PhD, PStat[1,2], Allan J. Kozlowski, PhD, PT [3], & Michael J. Strube, PhD[4]

[1]Program in Physical Therapy, Washington University School of Medicine

[2]Department of Neurology, Washington University School of Medicine

[3]Kozlowski Health Services Consulting LLC

[4]Department of Psychological and Brain Sciences, Washington University in Saint Louis

**Corresponding Author:**

Keith R. Lohse, PhD, PStat

Program in Physical Therapy, Department of Neurology

Washington University School of Medicine

E: lohse@wustl.edu

P: 314-286-1426

**Abstract**

Mixed-effect models are flexible tools for researchers in a myriad of fields, but that flexibility comes at the cost of complexity and if users are not careful in how their model is specified, they could be making faulty inferences from their data. We argue that there is significant confusion around appropriate random effects to be included in a model given the study design, with researchers generally being better at specifying the fixed effects of a model, which map onto to their research hypotheses. To that end, we present an instructive framework for evaluating the random effects of a model in three different situations: (1) longitudinal designs; (2) factorial repeated measures; and (3) when dealing with multiple sources of variance. We provide worked examples with open-access code and data in an online repository. We think this framework will be helpful for students and researchers who are new to mixed effect models, and to reviewers who may have to evaluate a novel model as part of their review.

Mixed-effect models (1–3) are an increasingly popular analytical method in a wide range of disciplines. As a crude illustration of its popularity, we conducted a literature search in PubMed to identify articles that reported using an Analysis of Variance (ANOVA) versus mixed-effect models. While ANOVA was by the far the dominantly used term, various references to mixed-effect models (sometimes also called mixed-effects regression, mixed models, or multi-level models(4,5)) have been closing the gap. This increase in the use of mixed-effect models is multi-causal, but broadly we think that researchers are attracted to mixed-effect models for their *flexibility*: mixed-effect models can deal with missing longitudinal data more easily; mixed-effect models highlight individual differences in the estimation of variance components; for longitudinal designs, mixed-effect models explicitly model individual trajectories and variability in time itself; and mixed-effect models allow for more than one source of variability to be estimated simultaneously(6–9). The mixed-effects framework also extends to other statistical applications such as meta-analysis(10), measurement (11), and mediation models (12,13). Finally, as with general linear models, mixed-effect models can be extended to handle categorical outcome measures.(14)

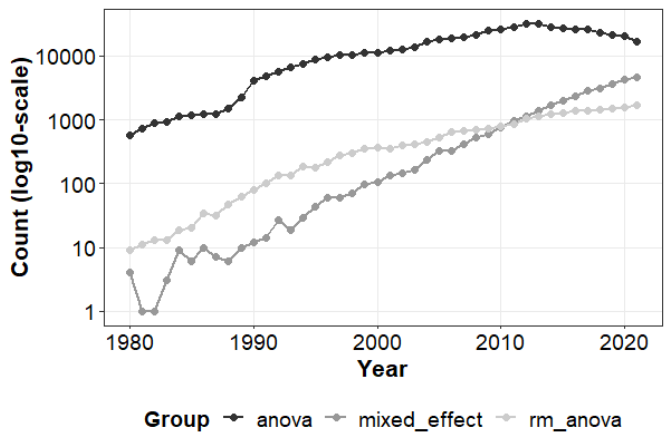

**Figure 1**. Frequency of hits on PubMed searched 2021-11-15 for "ANOVA", "mixed-effect/s" (or "mixed effect/s") and "repeated measures+ANOVA" (or "RM" + "ANOVA").

The tremendous power of mixed models comes at the cost of analytical complexity. Despite excellent resources on the topic (5,8,9), researchers are often faced with difficult choices and sometimes conflicting recommendations (15,16), and may not have the time or statistical expertise necessary to understand these resources (17), as many references are quite mathematically dense. Although there are domain-specific guides that are written to address particular types of research designs and use examples familiar to subject-matter experts in that domain (e.g., 18,19), confusion surrounds this complicated topic.

Our own personal experience, which includes a non-random sample of dissertation committees, manuscript reviews, and questions at workshops on mixed-effect models, shows that a major part of this confusion is in choosing the appropriate *random effects* for a model, with models often being mis-specified. Researchers generally appear better at selecting and interpreting appropriate *fixed effects* to test their hypotheses. These fixed effects correspond most often to the hypotheses that guided the research. However, the inferences about fixed effects depend on the random effects of the model.

Inappropriate specification of random effects has substantial negative consequences on the credibility of the statistical inference. For instance, including only a random effect of participant in a study with multiple within-subject factors can make it seem like the researchers have more independent data than they have, spuriously shrinking the standard error and increasing the Type I error rate (an example of *under*-specification). Conversely, including fully crossed random effects of participants and stimuli in an experiment where there is only one observation per person for each stimulus can leave no residual error, creating a non-identifiable matrix (an example of *over*-specification; (20)). Or the researchers may simply omit a random-effect that is justified either by the design or the empirical variability in the data (an example of *mis*-specification). Thus, reasonable specification of random effects is critical to making reasonable inferences about the fixed effects.

Our primary contention is that there is confusion among researchers about how to correctly specify mixed-effect models (as others have argued (21–23)). Our goal is thus to provide a technically sound but accessible and didactic paper to help researchers choose appropriate random effects as a starting point based on their study design. Our target audience is busy researchers who are not experts: such as, novices who are starting out with mixed-effect models, researchers who are familiar with the basic concept but unsure how to properly specify a model in a more complicated design, and reviewers tasked with evaluating a novel model in a manuscript review.

We will not be able to move someone from novice to expert with the concepts presented in this manuscript, but we can help researchers avoid some fundamental errors and prepare them to tackle more detailed texts (24–26). Our objective is to get researchers to think more broadly and carefully about random effects and the need to treat them appropriately in model specification. We also assume that all readers will be familiar with the language of "traditional" ordinary least squares regression (e.g., normality of residuals, contrast versus treatment coding, types of sums of squared errors, collinearity). If any of these terms are unfamiliar, we recommend studying those issues first (27–29). Mixed-effect regression is an extension of the general linear model and as such starting with simpler models is very helpful for tackling more complex designs.

Finally, we want to address some issues we will not cover. Our focus is on selecting a reasonable random-effects structure given the study design. As we will see this is more of a starting point and there is no universally correct structure (15,16), but there are some common incorrect structures that authors should avoid. Our manuscript is therefore not a general introduction to mixed-effect models (see (19,20)), nor we will discuss the intricacies of how one should specify fixed effects, interpret parameters in the presence of interactions, or select between competing models (see (8,9,24,25)). We will also not address advanced concepts like statistical power in mixed-effect regression (see (30,31)) nor the application of Bayesian estimation to mixed models (see (16)).

With that in mind, we first define notation and terms then address common designs as follows: 1. Longitudinal Designs, 2. Repeated Measures Factorial Designs, and 3. Designs with Multiple Sources of Variance. We try to keep our discussion of these issues concise here, but we provide vignettes with worked examples in an online repository (https://keithlohse.github.io/mixed_effects_models/).

**Notation and Terms**

In the manuscript, we focus on using `R` code(32) and specifically `lme4` and `nlme` syntax(26,33) to illustrate how models are specified. We provide `R` code in the text (and supplement) to facilitate implementation as `R` is free to use, and all code and data are also provided in our online repository. We focus on code to make the conceptual importance of random effects clearer to readers who may be less familiar with mathematical notation and matrix algebra. In the example below, we have a single within-subject variable (`W1`), so we need to account for within-subject clustering of the data. Including a random intercept for each subject (`1|subject`) removes the between-subjects variance from the outcome (`DV`), so that our independent variable is appropriately explaining within-subject variance after between-subjects variance has been removed.

**Code Snippet 1:**
```
# Note that R uses the '#' symbol to indicate programmer comments

example_model <- lmer(DV~
                      # Fixed-effects
                      1+ W1 +
                      # Random-effects
                      (1|subject),
                      data=DATA, REML=FALSE)

# Note that DV is the name of a continuous outcome, W1 is a factorial within-
# subject variable, and subject is a factorial subject identifier.
```

In the text, we indicate variables in a dataset with `Courier font`. For instance, we might generally speak of a "subject" participating in an experiment, but `subject` refers to a vector of subject identifiers (e.g., a column of "s1", "s2", as shown in Code Snippet 2). We also specify variables that are categorical factors versus those that are continuous predictors. For instance, when analyzing data from a randomized controlled trial, `condition` could be a vector of labels (e.g., "Experimental" versus "Control") whereas `time` could be a vector of times in days from enrollment (e.g., 0, 10, 35 days).

When presenting model outputs, we show F-tests that use Satterthwaite's approximation (34,35) to obtain degrees of freedom and p-values which will be of interest to most readers (but see (36–38)). We use Satterthwaite's approximation for its computational efficiency and to obtain degrees of freedom that we can compare to factorial ANOVA results (39). There are other methods available such as the Kenward-Roger approximation (40) or model comparison approaches (8,41), but a detailed discussion of all of these approaches is beyond our scope. Finally, we do not use report $r^2$ measures of effect-size as a direct $r^2$ calculation is not available in mixed-effect models (but see (42,43) for pseudo-$r^2$ measures).

**Defining Fixed and Random Effects**

There are different definitions of fixed and random effects in the literature (9,18,44,45), but our definitions are consistent with the frequentist mixed-effect estimation employed in `lme4` (26) and many domains of research (8,18). *Fixed effects* are those effects whose levels are fully represented in the data and beyond which we do not want to generalize. *Random effects* are those sampled from a larger population and from which we do want to generalize. For instance, in the code below, if we measured subjects' heart rate at rest and again following a bout of exercise, then `condition` (at rest or following exercise) would be a fixed effect. We are only interested in the difference between that specific type/duration of exercise and rest. In contrast, `subject` would be a random effect, because we have a random sample of subjects, and we want to generalize from that sample back to the larger population. Programmatically, that could be written:

**Code Snippet 2:**
```
example_mod <- lmer(heart_rate~
                    # Fixed-effects
                    1+ condition +
                    # Random-effects
                    (1|subject),
                    data=DATA, REML=FALSE)

# Based on the data below:
> DATA
  subject condition heart_rate
1      s1       ctl         60
2      s1        ex         72
3      s2       ctl         42
```

| 4 | s2 | ex | 50 |

Formally, the fixed and random effects in a regression model are written as a set of vectors and matrices:

(Eq. 1).    $y = X\beta + Z\gamma + \epsilon$

where **y** is the vector of scores on the dependent variable for each person at each time-point, **X** is the *design matrix* for the fixed effects, **β** is a vector of the fitted coefficients, **Z** is a design matrix for the random-effects, **γ** is a vector of fitted random deviates, and finally **ε** is a vector of random errors.

Although written in matrix notation, all we are really saying is that "the data = the model + the error" (41). However, our model has two components to it: the structural part, represented by the fixed-effects, **Xβ**, and the stochastic part, represented by the random-effects, **Zγ**.(4) In this heart rate example, there were only 2 subjects (S1 and S2) measured in each of the 2 conditions: during a controlled rest, "ctl", and following exercise, "ex".

As shown in Figure 2, we have one subject with a higher average heart rate (S1 = 60, 72) and one subject with a lower-than-average heart rate (S2 = 42, 50), in each of the conditions. Our random intercept of '`(1|subject)`' accounts for average between subject differences in **y** with the design matrix **Z**. This design matrix spreads out our random-effects, **γ**, into appropriate partitions for each subject; that is, every observation for S1 gets +9.9 and every observation for S2 gets -9.9. We can then add this result to the fixed effect estimates from the model. The fixed effect design matrix, **X**, indicates that there is a constant intercept (column of 1's) that gets further modified by a slope in every other instance (columns of 0's and 1's). The specific values for this intercept and slope come from the fixed-effects coefficients, **β**. Thus, our model boils down to a regression equation for each person (i.e., 51 + 10(*condition*), where condition = 0 is rest and = 1 is exercise) with a unique deviate added to observations coming from the same subject (i.e., -9.9 or 9.9). Critically, accounting for the statistical dependency between observations with the appropriate random effects allows us to get independent residuals for each observation (ε's) and thus appropriately test hypotheses about the fixed effects.

**A** $$y_{ij} = X\beta + ZU + \epsilon_{ij}$$

**B** $$\begin{bmatrix} y_{11} \\ y_{12} \\ y_{21} \\ y_{22} \end{bmatrix} = \begin{bmatrix} 1 & x_{11} \\ 1 & x_{12} \\ 1 & x_{21} \\ 1 & x_{22} \end{bmatrix} \begin{bmatrix} \beta_0 \\ \beta_1 \end{bmatrix} + \begin{bmatrix} 1 & 0 \\ 1 & 0 \\ 0 & 1 \\ 0 & 1 \end{bmatrix} \begin{bmatrix} U_1 \\ U_2 \end{bmatrix} + \begin{bmatrix} \epsilon_{11} \\ \epsilon_{12} \\ \epsilon_{21} \\ \epsilon_{22} \end{bmatrix}$$

**C** $$\begin{bmatrix} 60 \\ 72 \\ 42 \\ 50 \end{bmatrix} = \begin{bmatrix} 1 & 0 \\ 1 & 1 \\ 1 & 0 \\ 1 & 1 \end{bmatrix} \begin{bmatrix} 51 \\ 10 \end{bmatrix} + \begin{bmatrix} 1 & 0 \\ 1 & 0 \\ 0 & 1 \\ 0 & 1 \end{bmatrix} \begin{bmatrix} 9.9 \\ -9.9 \end{bmatrix} + \begin{bmatrix} -0.9 \\ 1.1 \\ 0.9 \\ -1.1 \end{bmatrix}$$

**D** $$\begin{bmatrix} 60 \\ 72 \\ 42 \\ 50 \end{bmatrix} = \begin{bmatrix} 51 + 0 \\ 51 + 10 \\ 51 + 0 \\ 51 + 10 \end{bmatrix} + \begin{bmatrix} 9.9 + 0 \\ 9.9 + 0 \\ 0 - 9.9 \\ 0 - 9.9 \end{bmatrix} + \begin{bmatrix} -0.9 \\ 1.1 \\ 0.9 \\ -1.1 \end{bmatrix}$$

**Figure 2**. Illustration of matrix notation for a mixed-effect model (A), which can be expanded into more detailed notation for our heart rate example with two observations and two subjects (B). Fitting the actual values from the data (C) allows us to combine the fixed effects and the random effects with their respective design matrices, to get one estimate from our model (fixed + random effects) for each subject in each condition (D). The difference between the model's prediction and the actual value ($y$'s) is the residual-error term ($\epsilon$'s).

**Nested and Crossed Random Effects**

Readers entering the world of mixed-effect models are likely to have heard of random-effects being either nested or crossed. *Nested* refers to situations where levels of one factor are only represented at one level of another factor. Such nesting occurs in fully hierarchical models.(4,5) For instance, children are nested within families, students might be nested within classrooms, and classrooms may be nested within schools. Note that nesting is a product of the research design/environment in which a study is conducted and may not be immediately apparent in how variables are coded. Figure 3A shows an example of *unambiguously* coded nested factors: `students` (S1-6) are nested within different `classrooms` (C1-3) with no student belonging to multiple classrooms. As seen in the cross tabulation, each student has only one observation in only one of the classrooms. Figure 3B shows an example of an *ambiguously* coded nested factor. The same two levels of student (S1-2) show up at each level of each level of classroom (C1-3). However, student S1 in class C1 is not the same student as S1 in C2 or S1 in C3. To make this clear, we can concatenate the classroom and student labels (`classroom:student`) to get a unique identifier for each student (e.g., C1:S2). The cross tabulation then shows the correct nesting: one observation from each student in only one class. If we did not concatenate these labels, we would end up with the wrong cross tabulation with 3-observations each for S1 and S2, one in each classroom, which would resemble the crossed random effects in Figure 2C.

*Crossed* refers to situations when different levels of one factor are represented at different levels of another factor. Two variables are said to be *fully crossed* if all levels of one factor are represented at all levels of the other factor. For instance, building on our heart rate example, let's say that both subjects (S1-2) had their heart rate measured in three conditions (C1-3: at rest, immediately following exercise, and 1-hr post exercise). If there are no missing data, we would get the fully crossed design shown in Figure 2C; each subject has three observations, one for each condition in which they were measured. If there are some missing data, these factors might become *partially crossed* as in Figure 2D (S1 is missing an observation for C3, S2 is missing an observation for C2).

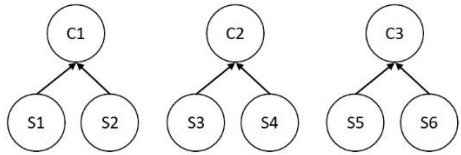
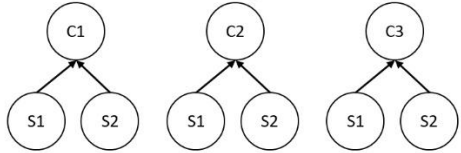
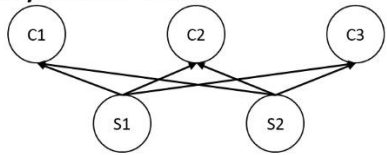
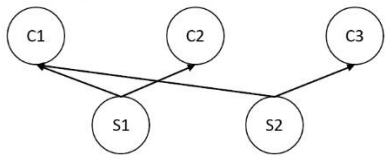

**Figure 3**. Schematic illustration of nested (A/B) and crossed (C/D) factors with their cross-tabulations on the right. C1-3 refer to levels of the categorical factor C. S1-2 refer to levels of the categorical factor S. In the nested case, we could think about S as students and C and classrooms (different students are in each classroom). In the crossed case, we could think about S as subjects and C as conditions in an experiment (every subject is measured in every condition).

Using `lme4` in `R`, researchers have several different options for inputting nested and crossed random effects as illustrated in Code Snippet 3:

**Code Snippet 3:**
```
# For students nested within classrooms we could use
(1|classroom) + (1|classroom:student)

# Or equivalent short cut using a slash "/":
(1|classroom/student)
# this syntax will work for ambiguous and unambiguous coding, see Figure 2

# If you have **unambiguously** coded nested factors, you can also use:
(1|classroom) + (1|student)
```

```
#For fully or partially crossed random effects, you would use:
(1|condition) + (1|subject)
```

If you have ambiguously coded nested factors, that means you would need to specify `(1|classroom) + (1|classroom:student)`. Thus, there is a random intercept for each classroom and for each unique student (`classroom:student` concatenates the levels of Classroom, n=3, and Student, n=2, making 6 unique labels). `lme4` also provides a shorthand way of representing this using the '/' operator, `(1|classroom/student)`. If you have unambiguously coded data, you could instead use `(1|classroom) + (1|student)` and get the same result. This code is equivalent because in unambiguous data, `student` has six levels (S1-S6) and in ambiguous data, the concatenation `classroom:student` also has six levels ('C1:S1' to 'C3:S2').

For fully and partially crossed random effects, one would use the syntax `(1|condition) + (1|subject)`. Note that this syntax has the same form as the unambiguously coded nested effects: `(1|classroom) + (1|student)`. This similarity helps to illustrate that the nesting (or crossing) of the random effect is not a property of the model, but a property of the experimental design. The model does not actually "know" which factors are crossed and which are nested, rather it depends on how those data are coded and selection of the appropriate syntax to suit the experimental design. Critically, a model will likely converge and produce an output of fitted coefficients with any of these random effects, and no warning that a user has the wrong random effects. This highlights the importance of researchers thinking critically about how their data are coded, then selecting the appropriate random-effects syntax.

# 1. Longitudinal Designs

**Random Slopes and Intercepts**

To illustrate some of the modeling concerns with longitudinal data, let us consider the hypothetical study depicted in Figure 4A. This plot shows data from three different groups of participants who had different types/levels of spinal cord injury (high tetraplegia, cervical vertebral levels 1-4; low tetraplegia, cervical vertebral levels 5-8; or paraplegia, vertebral levels thoracic 1- sacral 5) and reported their functioning in the activities of daily living at approximately monthly intervals for 1.5 years as they underwent physical and occupational therapy. Functioning is an umbrella term for how body structures/functions (e.g., level of spinal cord injury) interact with personal and environmental factors (e.g., accessibility), to shape activity limitations and participation restrictions.(46,47) Functioning in this case is the opposite of disability. For ease of interpretation, we constructed a fictitious measure of functioning on which scores range from 0 (very high disability) to 100 (very high function) and has interval/ratio properties. (Health science researchers might be more familiar with this construct as "independence", but we use "functioning" to avoid confusion with the statistical meaning of independence.)

As shown in Figure 4A, these data violate the assumptions of the ordinary least squares regression model in which errors are assumed to be independent and identically distributed. This violation can be seen most clearly when we consider that these residuals belong not only to each time point but are also unique to each participant. Consider the lowest scoring participant in the C1-4 group in the left panel of Figure 4A, regardless of the time point, this participant is going to have negative residuals (scoring lower than average relative to their group or to the sample as a whole). In contrast, a high scoring participant will have mostly positive residuals. This similarity of residuals within a participant reflects a correlation in the residuals (negatives with negatives, positives with positives), violating the assumption of independence. Violating this assumption is not trivial and can lead to considerable bias in the results of hypothesis tests.(48,49)

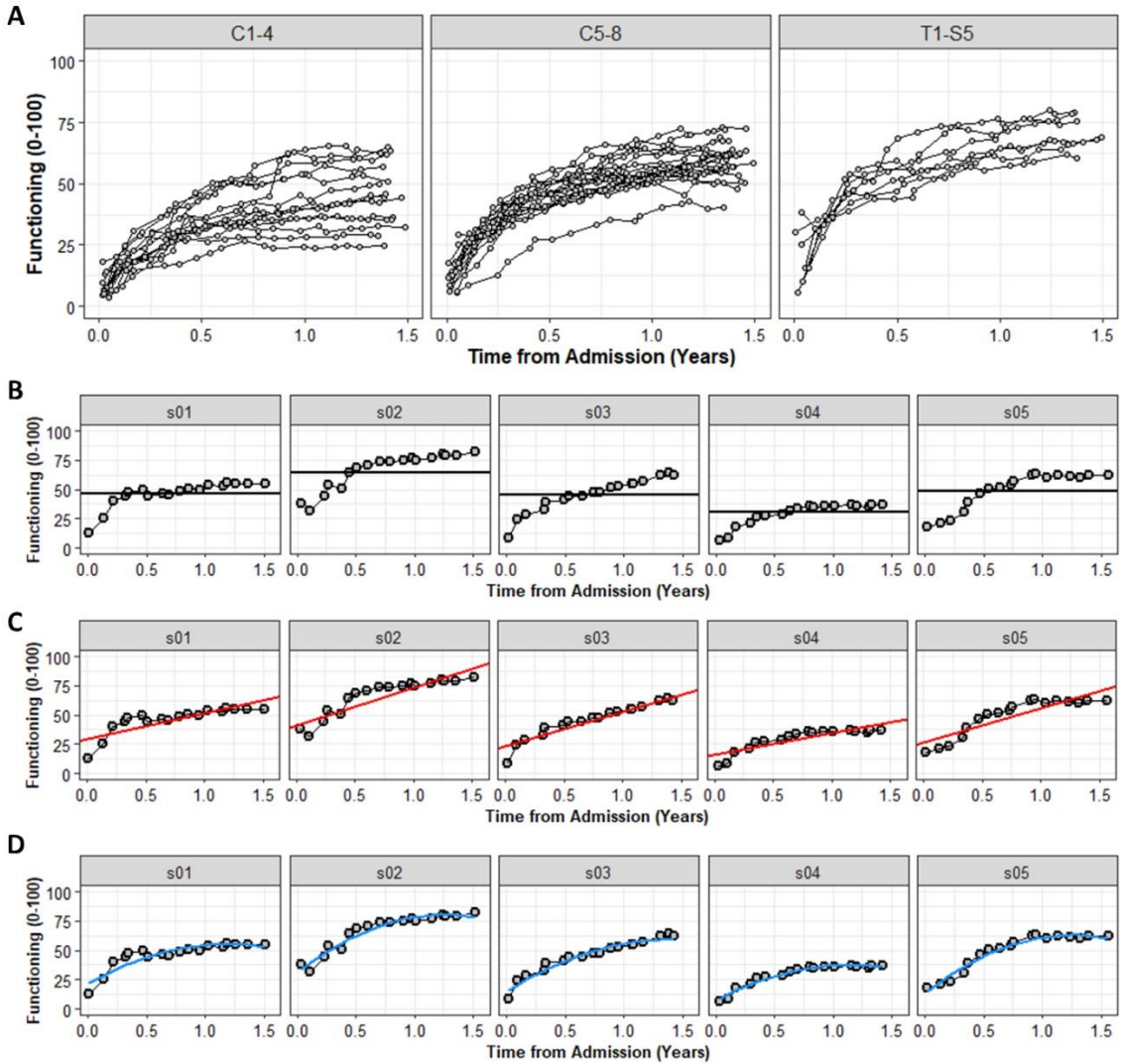

**Figure 4**. (A) Hypothetical functioning scores by time from admission for three different groups of participants with spinal cord injury. (B) Predicted values from a *random-intercepts* model (shown as black lines) plotted over the data for the first five participants. (C) Predicted values from a *linear random-slopes* model (shown as red lines) plotted over the data for the first five participants. (D) Predicted values from a *quadratic random-slopes model* (shown as blue lines) plotted over the data for the first five participants.

To account for this statistical dependence, the first step is to recognize that time points (`time`) are nested within subjects (`subject`) in this design. To do this, we can add a random intercept of subject to our model, as shown mathematically,

(Eq. 2A.)    $y_{ij} = \beta_0 + \gamma_{0i} + \epsilon_{ij}$

where:

(Eq. 2B.)　　　$\gamma_{0i} \sim N(0, \sigma_0^2)$

or programmatically:

**Code Snippet 4:**
```
raneff_int <- lmer(functioning~
                        # Fixed effects
                        1+
                        # Random effects
                        (1| subject),
                        data=DATA, REML=FALSE)
```

The predictions of this *random-intercepts model* are shown in Figure 4B for the first five subjects. Each subject gets their own unique intercept ($\beta_0 + \gamma_{0i}$), but there is considerable error not explained by the model ($\epsilon_{ij}$). By adding additional fixed and random effects to the model, we can reduce error at the level of the participant (*i*'s) and each time point (*j*'s).

To improve this model, we can estimate a unique trajectory for each subject by adding a fixed and random effect of `time`. This is often referred to as a *random-slopes* model, shown mathematically:

(Eq. 3A.)　　　$y_{ij} = \beta_0 + \beta_1(time_{ij}) + \gamma_{0i} + \gamma_{1i}(time_{ij}) + \epsilon_{ij}$

where:

(Eq. 3B.)　　　$\begin{bmatrix} \gamma_{0i} \\ \gamma_{1i} \end{bmatrix} \sim N\left(\begin{bmatrix} 0 \\ 0 \end{bmatrix}, \begin{bmatrix} \sigma_0^2 & \sigma_{01} \\ \sigma_{01} & \sigma_1^2 \end{bmatrix}\right)$

or programmatically:

**Code Snippet 5:**
```
raneff_lin_slope <- lmer(functioning~
                        # Fixed effects
                        1+time+
                        # Random effects
                        (1+time| subject),
                        data=DATA, REML=FALSE)
```

The predictions of this linear random-slopes model can be seen in Figure 4C. Now, not only does each subject have a unique intercept ($\beta_0 + \gamma_{0i}$), but also a unique slope [$(\beta_1 + \gamma_{1i})time_i$], which greatly reduces the amount of residual error ($\epsilon_{ij}$). Note also that one does not need add these fixed and random effects in one step (4,5,8). A fixed-effect could be added alone, which assumes that all participants changed at about the same rate, or a random-effect could be added alone, but that would assume the average effect of time is 0 across subjects. Importantly, having random-intercepts and random-slopes in the model also means that we must account for their covariance ($\sigma_{01}$ in Eq. 3B). In applied modeling, it is common to find that individual differences in slopes and intercepts will be correlated: either positively because individuals with higher intercepts will show greater change over time (i.e., "the rich get richer") or negatively correlated because individuals with lower intercepts will show greater change over time (i.e., "more room to grow" due to floor/ceiling effects).

Finally, we might want to account for the nonlinearity that we observe in the data. An intuitive way to account for this nonlinearity that is a direct extension of the linear model is to add a quadratic effect of time (i.e., `time`$^2$) to the model. Adding both a fixed- and random-quadratic effect of time yields:

(Eq. 4A.) $\quad y_{ij} = \beta_0 + \beta_1(time_{ij}) + \beta_2(time_{ij}^2) + \gamma_{0i} + \gamma_{1i}(time_{ij}) + \gamma_{2i}(time_{ij}^2) + \epsilon_{ij}$

where:

(Eq. 4B.) $\quad \begin{bmatrix} \gamma_{0i} \\ \gamma_{1i} \\ \gamma_{2i} \end{bmatrix} \sim N \begin{bmatrix} 0 \\ 0 \\ 0 \end{bmatrix}, \begin{bmatrix} \sigma_0^2 & \sigma_{01} & \sigma_{02} \\ \sigma_{01} & \sigma_1^2 & \sigma_{12} \\ \sigma_{02} & \sigma_{12} & \sigma_2^2 \end{bmatrix}$

or programmatically:

**Code Snippet 6:**
```
raneff_quad_slope <- lmer(functioning~
                          # Fixed effects
                          1+time+I(time^2)+
                          # Random effects
                          (1+time+I(time^2)|subject),
                          data=DATA, REML=FALSE)
```

In this model, as shown in Figure 4D, each participant not only has a unique starting point (the random intercept) and a unique initial slope at the intercept (the random linear slope), but also a unique curvature as the trajectory changes over time (the random quadratic slope).

Importantly, however, our model becomes much more complicated as we add random-slope parameters. Note that the addition of quadratic random-slope parameter did not merely add one parameter to our model, it actually added three: one for the variance of the quadratic slope ($\sigma_2^2$), one for the covariance of intercept and the quadratic slope ($\sigma_{02}$), and one for the covariance of the linear and quadratic slopes ($\sigma_{12}$), as shown in Eq. 4B. Thus, adding fixed- and random-effects of polynomials can improve the fit of our model, but we need to be thoughtful when doing so. Additionally, researchers need to consider the *centering* of the time variable when adding polynomials to the model. If time is centered on the first observation (i.e., `time` = 0 reflects the first observation) then higher polynomials will often be highly correlated, which can create convergence issues for the model. For detailed accounts of how to build these polynomial models, we refer readers to more thorough sources that explain how to interpret these effects, test the variance explained by each random-effect, and compare models with different random-effects in order to choose the most parsimonious model (4,8).

**Random-Effects in Non-Linear Models**

The quadratic random slopes model above is what we refer to as a *curvilinear model*. That is, the model ultimately predicts a nonlinear trajectory for each person, but the model itself is linear in its components: $\beta_0$ is simply added to $\beta_1 \cdot time_i$ and added to $\beta_2 \cdot time_i^2$ as in Eq. 4A. Forcing this nonlinearity via polynomials has strengths and weaknesses. Strengths include the ease of interpreting polynomials (e.g., the average reader can more easily conceive of `time`² compared to the log of `time` or `time` exponentiated) and they often correspond to hypotheses that researchers might have (e.g., the typical researcher might be interested in *any* evidence of curvature, rather than a specific power law or exponential law). However, a major weakness is that few biological or social systems follow a quadratic function (i.e., being u- or n-shaped and symmetric about a single inflection point). For instance, in Figure 4D, it is likely that participants were approaching an asymptote that could be better captured by a truly nonlinear model such as a negative exponential or logistic model.

A detailed discussion of nonlinear models lies outside of our focus on random effects, but see Chapter 6 of (4), and (25,33). To summarize however, (1) regardless of the form the nonlinear model takes, there is the possibility to include random effects for each parameter, and that (2) adding random effects can often lead to issues of over-fitting, so the random effects included in the final model should be justified through a combination of theoretical rationale and empirical model fit.

Here, we present an illustrative example of a negative exponential model. First, let us consider the general form of a negative exponential function:

(Eq. 5A). $\quad y_{ij} = \alpha_i + \delta_i e^{[(\lambda_i)Time_{ij}]} + \epsilon_{ij}$

there is an asymptote, $\alpha_i$, that defines the plateau for each person; there is a change parameter, $\delta_i$, that defines the y-axis location of a pseudo-intercept relative to the asymptote (intercept = $\alpha - \delta$); and there is a rate parameter, $\lambda_i$, which defines how quickly the participant will approach their asymptote. By convention, we write these parameter estimates as $\alpha_i$, $\delta_i$, and $\lambda_i$. However, these coefficients are not conceptually different from the $\beta$'s that we used in a linear model. Therefore, in the next equation, we will revert to $\beta$'s and $\gamma$'s, but otherwise keep the same negative exponential function:

(Eq. 5B). $\quad y_{ij} = [\beta_0 - \gamma_{0i}] + [\beta_1 - \gamma_{1i}]e^{[(\beta_2+\gamma_{2i})time_{ij}]} + \epsilon_{ij}$

Our model now consists of fixed effects, $\beta_0$, $\beta_1$, and $\beta_2$, which apply to the whole sample, shown in Figure 5A. However, we also have unique asymptotes ($\beta_0 + \gamma_{0i}$), pseudo-intercepts ($\beta_1 + \gamma_{1i}$), and rate parameters ($\beta_2 + \gamma_{2i}$) for each subject which are the sum of the fixed effect and the random deviate for that subject. These coefficients are shown for the first four subjects in Figure 5B.

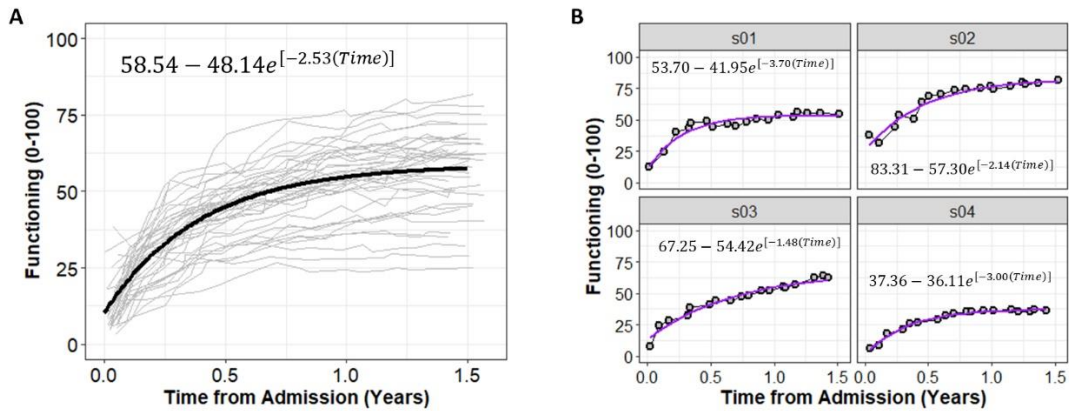

**Figure 5**. (A) Modeling the same hypothetical spinal cord injury data using a negative exponential function. Grey lines show the data for individual subjects. The black line shows the best fitting negative exponential model, with the fixed effects displayed in the panel. (B) Individual data (as points) and estimated trajectories (as purple lines) for the first four subjects. Equations show the coefficients for each participant (the group-level fixed effect from A, plus the random deviate for each participant).

Programmatically, we can also see how this model would be fit using `nlme`(33) in Code Snippet 7. A worked example of this non-linear model is provided in the online repository (as there are some important programming considerations for `nlme` versus `lme4`).

**Code Snippet 7:**
```
set.seed(100)
neg_exp_rand_mod <- nlme(functioning ~ b_0i + 
                          (b_1i)*(exp(b_2i * time)),
                    data = DATA,
                    fixed = b_0i + b_1i + b_2i ~ 1,
                    random = list(b_0i ~ 1, b_1i ~ 1, b_2i ~1),
                    groups = ~ subject,
                    start = c(80, -70, -1),
                    na.action = na.omit)
```

## 2. Repeated Measures Factorial Designs

In this section, we will consider situations in which we have categorical repeated measures, not continuous longitudinal data. For instance, a researcher might have a design with two within-subject factors of `condition` (conditions of rest, immediately following exercise, and following a 30-min delay) and `altitude` (laboratory simulated low- versus high-altitude environments). Although these measurements were taken at different times of day, a continuous measure of time is not what we want to model. In a well-controlled experiment, the factor should ideally be counter-balanced across participants to be independent of time.(50) Instead, we want to model the categorical difference between the two different sets of repeated measures: condition and altitude as shown in Figure 6A. Researchers who use these types of within-subject designs have historically analyzed their data using repeated-measures ANOVA, in which all factors vary within a subject. In our experience, many researchers want to apply the same conceptual analysis but utilize some of the strengths of a mixed-effect model (e.g., avoiding listwise deletion of missing data). But what is the appropriate random-effects structure that will make the model analogous to the repeated measures ANOVA? Consider five different options in Code Snippet 8:

**Code Snippet 8:**
```
# Head of the data file showing the first 12 observations (subject S1 and S2)
   subject altitude condition heart_rate
1       S1      low      rest         47
2       S1      low       imm         63
3       S1      low     delay         48
4       S1     high      rest         54
5       S1     high       imm         64
6       S1     high     delay         62
7       S2      low      rest         62
8       S2      low       imm         74
9       S2      low     delay         60
10      S2     high      rest         65
11      S2     high       imm         80
12      S2     high     delay         70
# …

# Option A: Random Intercept of Subject Only
heart_rate ~ 1 + condition*altitude +
             (1|subject)

# Option B: Random Intercepts for All Factors
heart_rate ~ 1 + condition*altitude +
```

```
                (1|subject)+
                (1|condition)+
                (1|altitude)

# Option C: Random Intercepts of Subject:Repeated Measures
heart_rate ~ 1 + condition*altitude +
              (1|subject) +
              (1|subject:condition) +
              (1|subject:altitude)

# Option D: Random slopes of condition and altitude within each subject
heart_rate ~ 1 + condition*altitude +
              (1+condition+altitude|subject)

# Option E: Random slopes including the interaction term
heart_rate ~ 1 + condition*altitude +
              (condition*altitude|subject)
```

When we analyze these data using a repeated measures ANOVA, we obtain main effects of Altitude, $F(1,9) = 58.36$, Condition, $F(2,18)=142.15$, and the Altitude × Condition interaction, $F(2,18)=8.63$. We can use these values as a benchmark to evaluate the performance of our mixed-effect models. Using these values as our "gold standard" assumes that the factorial ANOVA is the model that researchers want given their design. We acknowledge that this may not always be the case, but we think it is generally best to choose random effects based on the study design (as others have argued (51)) and given that most researchers would analyze data from this study with a two-way repeated-measures ANOVA, it is reasonable to use that benchmark.

Before we can evaluate these different random effects structures, it is important to briefly explain two different methods of estimation: full Maximum Likelihood (ML) and Restricted Maximum Likelihood (REML). Without getting into the details of how these estimators actually work (see (4,25,26)) a key difference is that ML will base estimates on all of the available information (i.e., fixed and random effects simultaneously) but does this as an iterative procedure. The fixed components are estimated first, and then random components estimated second (e.g., as the sample mean needs to be calculated to obtain the sample variance). In contrast, REML corrects for the degrees of freedom lost in estimating the fixed effects, but to do this REML invokes a different process in which variance components are estimated first and fixed effects are treated as a "nuisance parameter" (for an accessible discussion see (52)).

These computational differences mean that REML and ML have complementary strengths and weakness. For instance, if you wanted to compare models with different fixed effects to ascertain the best fitting model, then you need to use ML because it is agnostic to the degrees of freedom lost to the different fixed effects in the different models. However, ML is also prone to under-estimate the variance components, especially in small samples. Thus, if there is no model comparison (e.g., you are a fitting a single mixed model with random-effects determined by the design) or you are comparing models with identical fixed effects (e.g., trying to determine only the preferred random-effect structure), then REML will lead to more accurate estimation of the variance components than ML.

In practice however, this bias in REML versus ML should only substantially effect the estimation of the variance when the number of observations, $n$, is small relative to the parameters being estimated, $p$. Readers are likely familiar with this issue when calculating the standard deviation of a sample, where we divide the sum of squared errors by $n-1$ (analogous to REML) rather than simply $n$ (analogous to ML). Thus, the de-biasing effect of dividing by $n-p$ relative to just dividing by $n$ depends on the values of both $n$ and $p$. When the number of observations is much larger than the number of parameters ($n>>p$), then the choice between REML and ML will usually have a trivial effect on the variance components.

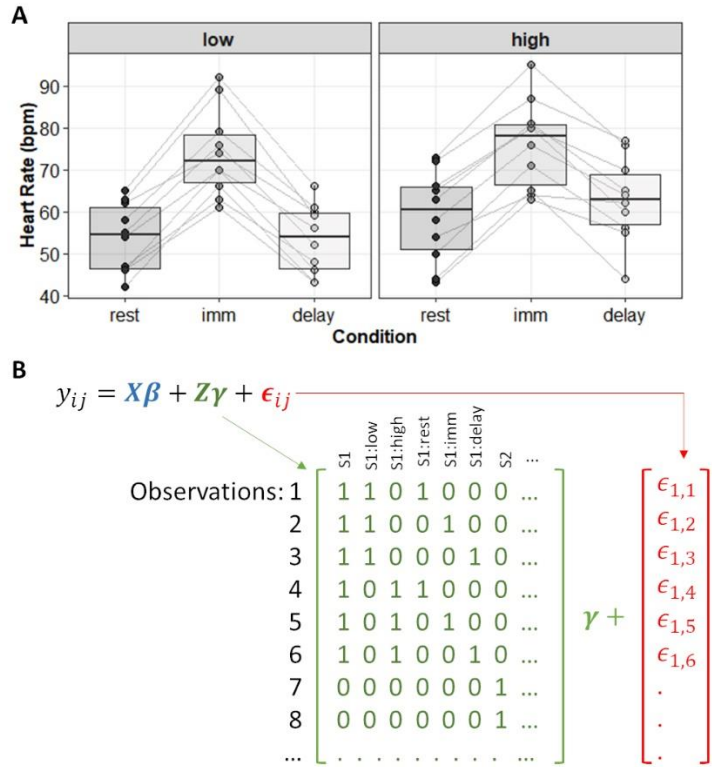

**Figure 6.** (A) Data from our hypothetical experiment in which the same subjects had their heart rate measured at simulated low and high altitude in the lab at three different time points: at rest, immediately following exercise, and following a 1-hr delay post exercise. (B) A truncated design matrix for the random-effects structure of the model. The first 8 observations (6 belonging to subject S1, 2 to subject S2) are shown in rows and the 'Subject:Factor' intercepts from the mixed model are shown in columns.

**Option A: Random Intercept of Subject Only – A Poor Choice**

In our experience, many researchers select Option A, including only a random intercept of subject, (1|subject). Including a random intercept for each subject is *necessary but not sufficient* to account for all the statistical dependencies in this within-subject design. This can be seen in Table 1, where we contrast the denominator degrees of freedom and F-values from a mixed-factorial ANOVA(39) and the five different mixed-effect models. Option A leads to greatly inflated denominator degrees of freedom for each of the effects; the denominator degrees of freedom are all 45, much larger than the

corresponding 9 or 18 degrees of freedom in the repeated measures ANOVA. This spuriously increases statistical power (note the larger F-values for each test), making it seem as though we have more independent pieces of data than we do. Inspecting the degrees of freedom helps to quantify the conceptual problem with this random-effects structure: we do not have 45 independent pieces of information and we need to account for the fact that data at each altitude are measured in each condition, in addition to being measured within each subject.

*Table 1.* Summary model output for a factorial repeated measures ANOVA and mixed-effects models using five different random-effects and two different methods of estimation (ML versus REML).

|  | **Main effect of Altitude** | | **Main Effect of Condition** | | **Altitude x Condition Interaction** | |
|---|---|---|---|---|---|---|
| **Model** | **(df1, df2)** | **F-Value** | **(df1, df2)** | **F-Value** | **(df1, df2)** | **F-Value** |
| ***Repeated Measures ANOVA*** | | | | | | |
| ANOVA, Type III sums of squares | **(1, 9)** | **58.36** | **(2, 18)** | **142.15** | **(2, 18)** | **8.63** |
| ***Random Effects - REML*** | | | | | | |
| Option A | (1, 45) | 43.13 | (2, 45) | 199.25 | (2, 45) | 6.29 |
| Option B | (1, 45) | 0.50 | (2, 45) | 1.52 | (2, 45) | 6.29 |
| **Option C** | **(1, 9)** | **58.35** | **(2, 18)** | **142.15** | **(2, 18)** | **8.63** |
| Option D* | (1, 12) | 43.88 | (2, 9) | 123.51 | (2, 27) | 10.51 |
| Option E† | na | na | na | na | na | na |
| ***Random Effects - ML*** | | | | | | |
| Option A | (1, 50) | 47.92 | (2, 50) | 221.39 | (2, 50) | 6.99 |
| Option B ‡ | (1, 50) | 47.92 | (2, 50) | 221.39 | (2, 50) | 6.99 |
| Option C | (1, 10) | 64.84 | (2, 20) | 157.95 | (2, 20) | 9.59 |
| Option D* | (1, 13) | 48.77 | (2, 10) | 137.26 | (2, 30) | 11.68 |
| Option E† | na | na | na | na | na | na |

*Note*. Readers can see all the underlying code, including the standard ANOVA functions in the online repository. Df1 refers to the numerator degrees of freedom and df2 refers to the denominator degrees of freedom for the calculation of the F-statistic. REML = Restricted maximum likelihood estimation; ML = full Maximum Likelihood estimation.
\* = returns warning "boundary (singular) fit: see help('isSingular')", correlations between random-effects approach 1, but there is no zero variance for every random effect.
† = Leads to an error, "Error: number of observations (=60) <= number of random effects (=60) for term (condition * altitude | subject); the random-effects parameters and the residual variance (or scale parameter) are probably unidentifiable". Without replicates, this model is over-specified and cannot be fit to these data.
‡ = returns warning "boundary (singular) fit: see help('isSingular')", the variance for (1|altitude) and (1|condition) both go to zero, effectively leaving this the same as Option A when fit using full maximum likelihood.

**Option B: Random Intercepts for All Factors – A Poor Choice!**

To account for the fact that observations are clustered within subjects, altitudes, and conditions, we have often seen authors use a random-effects structure like Option B. At first, these three random intercepts might seem reasonable, but there is a major conceptual flaw with this model: Our observations are not crossed by altitude and condition *overall*, they are only crossed by altitude and condition *within each subject*. The correct way to handle this is by nesting altitude and condition within each subject and accounting for those crossed factors at the within-subject level, which is precisely what we do in Option C. Before discussing that structure, however, let us dwell on Option B as an example of what not to do.

When estimated with REML, the main-effect for altitude and condition all but disappear with $F(1, 45)=0.50$ and $F(2,45)=1.52$, respectively. These F-values are greatly reduced because we have pulled out the variance due to `altitude` and `condition` in our random intercepts prior to the estimation of the fixed effects, leaving a trivial amount of variation to be explained. The interaction term remains similar to what we saw with Option A, $F(2,45)=6.29$, because that variance is not accounted for by the random intercepts.

When estimated with ML, the results of Option B are identical to Option A because now the fixed effects are fit first and the redundancy in the fixed effects now leaves no variance to be explained by the random intercepts of altitude and condition. The variance for both terms is estimated to be 0 (see footnote ‡ in Table 1), and we are effectively left with a random intercept of `subject`.

**Option C: Random Intercepts for Subject:Repeated Measures – A Good Choice!**

Random effects Option C is a very good choice to appropriately account for the statistical dependencies in these data. In that model, the degrees of freedom and the F-values match the repeated measures ANOVA exactly using REML. These outputs match because not only are we accounting for the fact that data-points come from the same subject, `(1|subject)`, but we also account for the fact that each subject is measured multiple times at each altitude, `(1|subject:altitude)`, and each

condition, `(1|subject:condition)`. As shown in a truncated design matrix in Figure 6B, having a unique intercept for each subject:factor accounts for the fact that altitude and condition are crossed factors, but only within each subject. As shown in Table 1, if we use ML instead of REML the results no longer match the ANOVA precisely, with degrees of freedom and F-values slightly increased. However, if we fit these models in a larger data set, the difference between REML and ML would be smaller.

It is important to note that this random-intercepts method gets more complicated with more within-subject factors. For instance, if we were to add a third within-subject factor to a fully factorial design (call it `W3`), then not only would this add an additional fixed effect, but it adds considerable statistical dependency that needs to be accounted for in the random effects. In that situation, observations come not only from each subject in each condition, but from each subject in each combination of the conditions, so we would need to add intercepts to account for the two-way interactions within each subject (e.g., `(1|subject:condition:altitude) + (1|subject:altitude:W3) + (1|subject:condition:W3)`). It is also important to note that our heuristic for the degrees of freedom may start to breakdown in more complex designs and although the degrees of freedom will be similar to the ANOVA, they will no longer match exactly (e.g., with sparse random-effects when there may only be two-levels of a factor). To save on space, we present a detailed example of a three-way within-subject design in the online supplement.

**Option D: Random Slopes of Condition and Altitude within in Subject – Also a Good Choice!**

Another reasonable alternative would be to include random slopes for each factor within each subject as in Option D. Conceptually, this model accounts for the within-subjects nature of the design similar to Option C, but it approaches the problem in a very different way and with more complexity that may not always be desirable. First, recall that when fitting random slopes, we do not only estimate their variance, but also their covariance. This allows us to account for potential correlations between the effect of altitude and condition across subjects, which is good, but increases the number of parameters being estimated, which is potentially problematic. To simplify the model, one could also constrain the random slopes to be independent (e.g., using "||" in place of "|" in `lme4` syntax). However, one would still run into a second issue which is that random slopes are actually a series of contrasts for categorical factors. That is, for `altitude` we only have two levels and therefore we need only one random slope to capture that difference. For `condition` we have three levels and therefore two random slopes are needed, which would also add more covariances if the model is not constrained. As the number of levels of the categorical factor/s increases, the model becomes substantially more complicated. Third, and finally, when using this random-slopes approach, the user needs to cognizant of what kind of contrasts are being used (as with centering of continuous variables). By default, `R` will use treatment coding with the first level, alpha-numerically, treated as the reference group, but this may not always be desirable and will affect the estimates and interpretation of the model.

**Option E: Random slopes including the interaction term – Only if you have replicates!**

Estimating the variance in our effects is a very useful step towards understanding individual differences (e.g., how much does the effect of altitude differ between people?), so the random slopes structure in Option D is appealing. Ideally, we would like to take this a step further and get an estimate of how the interaction between altitude and condition differs between people as well, so we can add the random slopes that capture that interaction in Option E. However, we cannot actually estimate this model given the current data. With only one observation for each subject in each condition, if we try to fit a unique three-way interaction for every subject, we will have perfect prediction and no residual error. This can be seen in the † footnote under Table 1, we only have 60 observations and if we want random slopes for each factor and their interaction, then we end up with 60 random effects. As the error message says, "Without replicates, this model is over-specified and cannot be fit to these data".

*Replicates*, as the name implies, are repeated observations at the same level of the data. For instance, assume that rather than one average heart rate for each subject in each condition, we ran our subjects through our exercise protocol on two different occasions with a sufficient washout period in between exposures to reduce order effects. We would now have two observations for each subject in each combination of altitude and condition, yielding 120 observations. Only two observations per person may not be enough to get very stable estimates of the variance for these within-subject effects, but it would give us the observations necessary to estimate a model using Option C. Thus, having replicates is often very desirable but comes with increased complexity and often implicit assumptions of which an author might not be aware. For instance, let's be generous and say that I have 10 observations for every combination of altitude and exercise. I could estimate a model using Option C, but that would assume that all 10 observations are independent of each other. Depending on how those observations were obtained, independence might not be a tenable assumption. For instance, there are likely temporal order effects and neighboring observations are more likely to be similar. As a result, our model might need to become more complex, for instance estimating a covariance structure for the residuals rather than assuming they are independent. (25,33)

### 3. Designs with Multiple Sources of Variance

A major strength of mixed-effect models over traditional ANOVA is their ability to handle different sources of variance simultaneously. We can see these different sources of variance when we have multiple factors that are random samples. For instance, this can happen when we have multiple levels of nesting, such as time within patients and patients within hospitals. That is, we want to generalize from our sample of patients and our sample of hospitals back out to a theoretical population of patients and population of hospitals, but we also need to account for the statistical dependence of patients from the same hospital. This sort of hierarchical nesting is why many mixed-effect models are referred to as "multi-level models" (MLM's (9)) or "hierarchical linear models" (HLM's (5)).

In the *Variation in Multiple Nested Factors* section below, we illustrate this kind of hierarchical nesting. We use the same hypothetical spinal cord injury data from the previous example of longitudinal data analysis. However, we now pretend that these patients were recruited and assessed at different study sites; so, we have time within subjects and subjects within sites. See Figure 7A.

We can also see different sources of variation when participants are repeatedly exposed to stimuli. In psychiatry, this might be a series of emotionally laden images that the participant is viewing as we measure the participant's emotional reactivity. The picture stimuli could be considered a random sample of all possible emotional images, and therefore we could include a random intercept for `image` as well as for `subject`. In speech pathology and audiology, the stimuli might be a series of words that the participant is hearing or saying. The words are a random sample of possible words and therefore we want to account for the variance in our sample of words in the same way we account for the variance in our sample of participants.

In the *Variation in Multiple Crossed Factors* section below, we will use a linguistic example from speech recognition (53). These response time data have been modified from the original data for didactic purposes (19) and illustrate how we can account for variation due to both our sample of subjects and the sample of words. See Figure 7B/C.

In that experiment, participants listened to words in an audio only modality in which a speaker was heard but not seen, and in an audiovisual modality in which the speaker could both seen and heard. Across all participants some words were presented in both modalities, but each word was presented in only one modality for a given speaker. The dependent variable was participants' response time in a second task in which they had to decide if a tactile stimulus was presented for a short, moderate, or long duration. The response on this secondary task is thus an index of the cognitive demand of processing the word. If the response time is faster, we can infer that the primary task required fewer cognitive resources. If the response time is slower, we can infer that the primary task required more cognitive resources. The researchers hypothesized that participants would be slower to respond in the audiovisual condition because participants would be processing both the audio and visual information available.

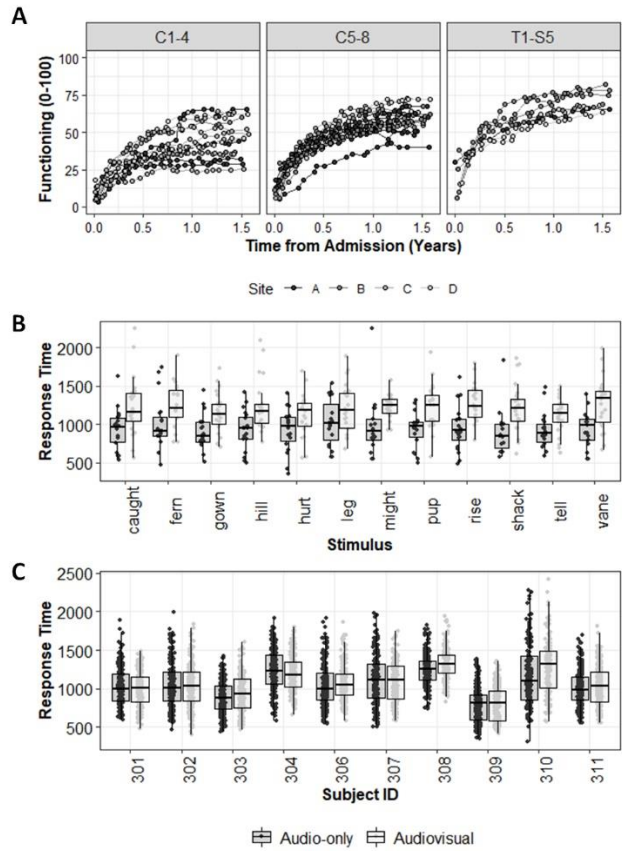

**Figure 7.** (A) Data from the hypothetical spinal cord injury study. Functioning in daily life is shown by time since admission and the study site at which the data were collected. Different study sites are shown in the fill colors, lines connect data points for individual subjects. (B/C) Data from the adapted linguistics example. Panel B shows response times as a function of the first ten words in the stimulus list (out of 543 total words). Points show the response time for individual subjects. Panel C shows response times as a function of the first ten subjects (out of 53 total subjects). Points show response times for individual words. In both B and C, points are color coded by the modality in which they were presented audio only, or in an audiovisual modality in which the speaker was seen.

**Variation in Multiple Nested Factors (e.g., Subjects within Study Sites)**

Returning to our hypothetical spinal cord injury data, in Figure 7A we can see multiple sources of statistical dependency. Time points are nested within individual subjects and neighboring time points are more likely to be similar than later time-points, but now we also have subjects nested within different research sites (A-D). We can account for this additional level of nesting in our R code by using a random intercept for study site:

**Code Snippet 9:**
```
raneff_quad_site<-lmer(function~
                            # Fixed-effects
                            1+time*AIS_grade+I(time^2)+AIS_grade+
                            # Random-effects
                            (1+time|subject)+(1|site), data=DAT1, REML=FALSE)
```

Fitting this model yields the random and fixed effects shown in Table 2, and yields the warning message "`boundary (singular) fit: see ?isSingular`". Looking at the random-effects output in Table 2 we can see why: the variance for the `site` random-effect is estimated to be zero, which is a "boundary" because variances can only be between $[0, +\infty)$. Correlations are similarly bounded between $[-1, +1]$. In this case, it would be reasonable to drop the site random effect from the model, as there is no variance being reliably explained by the different study sites. We could then proceed with a simpler random-effects structure.

*Table 2.* Summary of the model output for spinal cord injury data, including a random intercept for the different study sites.

| Random Effects | | | | |
|---|---|---|---|---|
| **Group** | **Name** | **Variance** | **SD** | **Correlation** |
| Subject | Intercept | 21.35 | 4.62 | |
| | Time | 48.01 | 6.93 | -0.08 |
| Site | Intercept | 0.00 | 0.00 | |
| Residual | | 11.83 | 3.44 | |
| **Fixed Effects** | | | | |
| **Effect** | | **F-Value** | **df1, df2** | **p-value** |
| AIS Grade | | 50.73 | 2, 50.7 | <0.001 |
| Time | | 1450.21 | 1, 140.4 | <0.001 |
| Time $^2$ | | 1104.46 | 1, 640.4 | <0.001 |
| Time x AIS Grade | | 3.83 | 2, 141.1 | 0.024 |
| Time$^2$ x AIS Grade | | 1.31 | 2, 640.5 | 0.271 |

* Note that the number of observations = 720, subjects = 40, and sites = 4.

However, there are a few considerations for dropping this random effect from the model. First, we only have four study sites in this example, which is very small number for the estimation of a random-effect (54,55). If we still wanted to account for differences between sites, we could include site as a fixed effect. If we include it as a fixed effect, we again see that there is very little variance explained by study site, $F(3, 40.0)=0.97$, $p=0.414$. So, in a practical sense, we can account for the variation due to study site in either format. Philosophically though, these two approaches are different. Treating study site as a fixed effect assumes that these are the only study sites in which we are interested. Treating study site as a random effect assumes sites are a random sample of possible sites. In either case, we should think about this lack of explained variance as a null result. That is, we have not proven that there is 0 variation between study sites, we have simply failed to find evidence of a difference between study sites – and the absence of evidence is not evidence of absence (56).

**Variation in Multiple Crossed Factors (e.g., Variation in Subjects and Stimuli)**

In our linguistic example, we must account for the fact we have multiple observations not only for each participant, but also for each stimulus. Note that unlike the longitudinal data, there is no reliable pattern to the data within a participant, because the order of the stimuli was randomized for each person. As such, we can account for this statistical dependence with random intercepts of `subject` and `stimulus`, and a fixed effect of the modality of the stimuli (`modality`).

**Code Snippet 10:**
```
rand01 <- lmer(log(RT)~
               # Fixed Effects
               1+modality+
               # Random Effects
               (1|subject)+(1|stimulus),
            data=DAT2, REML=TRUE)
```

As shown in Table 3, there was relatively less variability due to stimuli (SD = 0.017) compared to participants (SD=0.152). Accounting for these sources of variation, there was a statistically significant effect of modality, such that participants were generally slower to respond to the audiovisual stimuli compared to the audio-only stimuli, $b=0.079$, $F(1,21440)=541.48$, $p<0.001$.

*Table 3.* Summary of the model output for the linguistic data, including random intercepts of both participant and stimulus.

| Random Effects | | | |
|---|---|---|---|
| Group | Name | Variance | SD |
| Stimulus | Intercept | <0.001 | 0.017 |
| Subject | Intercept | 0.023 | 0.152 |
| Residual | | 0.061 | 0.248 |
| **Fixed Effects** | | | |
| Effect | F-Value | df1, df2 | p-value |
| Modality | 541.48 | 1, 21440 | <0.001 |

* Note that the number of observations = 21,679, stimuli = 543, and subjects = 53.

However, we should take note of the denominator degrees of freedom for this F-test, which is 21,440. This makes sense give the large number of observations: 543 stimuli x 53 participants + missing data = 21,679 observations. However, this is much larger than what we would expect from a comparable ANOVA design if modality had been included as a within-subject factor: $df_2 = (N-1)(k-1) = (53-1)(2-1) = 52$. The reason that these degrees of freedom are so much larger is because we have numerous replicates within each subject. That is, for each subject in each modality, we have multiple observations from several different stimuli. In this situation, we have a few options. We could aggregate the data down to one observation for each subject in each modality. This would make our mixed model more like a traditional repeated measures ANOVA with a within-subject factor of modality. Indeed, if we averaged over stimuli and then ran the code below, we would obtain a main effect of modality with F(1,52)=45.94, p<0.001.

**Code Snippet 11:**
```
DAT3 <- DAT2 %>% group_by(subject, modality) %>%
  summarize(RT = mean(RT))

rand_slopes <- lmer(log(RT)~
                # Fixed Effects
                1+modality+
                # Random Effects
                (1|subject),
              data=DAT3, REML=TRUE)
```

However, aggregating over stimuli would remove all the variability due to the individual stimuli. Thus, if we want to estimate the variability due to the specific stimuli we chose, we need to retain the different observations for the different stimuli. Instead of aggregating, we could retain the random intercept of `stimulus`, and add a random slope for the `modality` within each subject as in Code Snippet 12.

**Code Snippet 12:**
```
rand_slopes <- lmer(log(RT)~
                # Fixed Effects
                1+modality+
                # Random Effects
                (1+modality|subject)+(1|stimulus),
              data=DAT2, REML=TRUE)
```

We can see the results we obtain from this model in Table 4. By including a random intercept for stimuli and a random slope for modality within each subject, we can effectively estimate the variance due to the different stimuli (SD = 0.075) while reducing the degrees of freedom in the F-test, $F(1, 51.9)=52.86$, $p<0.001$. Conceptually, the reason the denominator degrees of freedom have shrunk from 21,440 to 51.9 is because we are no longer treating the modality effect as being estimated from 21,679 independent data points, but from 53 independent participants. Thus, if the model converges, including the random slope is going to be more appropriate when we have replicates in the data.

*Table 4.* Summary of the model output for the linguistic data, including random intercepts of both participant and stimulus, and a random effect of modality within each participant.

| **Random Effects** | | | | |
|---|---|---|---|---|
| **Group** | **Name** | **Variance** | **SD** | **Correlation** |
| Stimulus | Intercept | <0.001 | 0.017 | |
| Subject | Intercept | 0.025 | 0.159 | |
| | Modality | 0.006 | 0.075 | -0.29 |
| Residual | | 0.060 | 0.245 | |
| **Fixed Effects** | | | | |
| **Effect** | | **F-Value** | **df1, df2** | **p-value** |
| Modality | | 52.86 | 1, 51.9 | <0.001 |

* Note that the number of observations = 21,679, stimuli = 543, and subjects = 53.

**Conclusions**

First, we would like to thank readers for sticking with us and making it through a rather lengthy monograph. Second, we hope that this gives readers a solid starting point when thinking about what random-effects to include in their mixed-effect model given their study design. Longitudinal designs, factorial repeated measures, and studies with multiple sources of variance all have unique considerations for how to appropriately deal with statistical dependency in the data. We encourage practitioners to work through the vignettes provided in the online repository to get a better practical understanding of how to build these models in R and how to interpret the results. And finally, we hope readers have an appreciation for the complexity one faces when using mixed-effect models. The nuances of these modeling choices are not trivial and can have substantial effects on one's statistical conclusions.